\documentstyle[12pt]{article}
\begin{document}
\begin{center}
\begin{Large}
{\bf Diffractive} $\rho$-{\bf meson Leptoproduction \\
on Polarized Nucleon}\\
\end{Large}
\vspace{1cm}
\begin{large}
M. G. Ryskin\\
\end{large}
Petersburg Nuclear Physics Institute\\
188350, Gatchina, St.Petersburg, Russia
\end{center}
\vspace{2cm}
\begin{abstract}
The amplitude of diffractive $\rho$ (and open quark) leptoproduction
on a polarized target is calculated in the leading log approximation of
pQCD using the hadron-parton duality hypothesis. 
The spin-spin asymmetry is expressed in terms of the
 spin dependent gluon and quark structure functions  in the small $x$ region.
 Therefore the $\gamma^*+p\to \rho +p$ reaction provides 
a promising tool
to study the spin dependent gluon distribution $\Delta G(x,q^2)$.
\end{abstract}
\vspace{3cm}
E-mail: $\;\;\;\;$ ryskin@thd.pnpi.spb.ru
\newpage
\section{Introduction}
In ref.~\cite{MRT} it was demonstrated that  based on the hadron-parton
 duality one can discribe rather well the diffractive $\rho$-meson 
 electroproduction on the unpolarized target 
in the framework of perturbative QCD, not only for
the longitudinal part of the cros section but also for the transverse one .
 At large energies (i.e. in the small $x$ region) the amplitude of this process 
is proportional to the gluon  density and the cross section is 
proportional to the gluon  density squared.
 
Correspondingly, on the polarized nucleon the spin dependet part of the
 amplitude should be given in terms of the polarized parton densities 
$\Delta G$ and $\Delta q$. In the present paper, the same hadron-parton duality 
hypothesis will be used to calculate the spin dependent part of the 
amplitude of diffractive $\rho$ leptoproduction within the Leading Log 
 Approximation (LLA) of perturbative QCD.\footnote{An analogous calculatuion for
 $J/\Psi$ was done in \cite{JGN} assuming a nonrelativistic $J/\Psi$ 
meson wave function.}

First we recall the hadron-parton duality hypothesis. Let us consider the 
amplitude of the open $q\bar{q}$ leptoproduction, project it on the 
state with the spin $J^{PC}=1^{--}$ (isospin $I^G=1^+$) and average over a mass 
interval $\Delta M_{q\bar{q}}$ (typically $\sim 1$ GeV$^2$) in the region of 
$\rho$ mass. In this domain the more complicated partonic states 
($q\bar{q}+g,\; q\bar{q}+2g,\; q\bar{q}+q\bar{q},...$) are suppressed, while on 
the hadronic side the $2\pi$ states 
 are known to dominate. Thus for low $M^2$ we 
mainly have $\gamma^*\to q\bar{q}\to 2\pi$ or in other words
\begin{equation}
\sigma(\gamma^*p\to\rho p)\simeq \sum_{q=u,d}\int^{M^2_b}_{M^2_a}
\frac{\sigma(\gamma^*p\to (q\bar{q})' p)}{dM^2}dM^2
\end{equation}
where the limits $M^2_a$ and $M^2_b$ are chosen so that they appropriately
 embrace the $\rho$-meson mass region and $(q\bar{q})'$ means the projection of
 the $q\bar{q}$ system onto the $J^{PC}=1^{--}$, $I^G=1^+$ state.

The procedure of the projection onto the $J^{PC}=1^{--}$ state is described
 in sect. 2. Then in sect.3 we  calculate
 the spin dependent Born amplitude for $\gamma^*p\to q\bar{q} p$ process in the
 small $x=(Q^2+m^2_\rho)/s\ll 1$ region; here $Q^2=-q^2$ is the heavy photon 
virtuality and $s=(q+p)^2=W^2$ is the photon proton energy squared. Taking into 
account the leading log$Q^2$ corrections within the LLA in sect.4 we express 
the result in terms of the parton distributions $\Delta G$ and $\Delta q$ and
compare it with the formulae for the unpolarized case.  This way the spin-spin 
 asymmetry for the diffractive open quark production is calculated. Finally in 
sect. 5 we discuss the asymmetry for $\rho$-meson production.

\section{Projection onto the $J^{PC}=1^{--}$ state}

The cross section of diffractive open quark leptoproduction
\begin{equation}
\frac{d\sigma(\gamma^*p\to(q\bar{q})p)}{dtdM^2}=4\pi e^2_qN_c
\sum_\lambda\int\frac{d^2k_tdz}{16\pi^3}
\delta\left(M^2-\frac{k^2_t+m^2_q}{z(1-z)}\right)
\frac{|\mbox{M}_\lambda|^2}{16\pi s^2}\; ,
\end{equation}
where: $e_q$ is the electric charge of the quark (say, 
$e_u=\frac 23\sqrt{\frac{1}{137}}$), $N_c=3$ is the number of colours while 
$k'_t$ and $z$ are the transverse momentum and photon 
momentum fraction carried by the quark; $m_q$ is the quark mass, 
$\lambda$ is the quark helicity and 
$\mbox{M}_\lambda$ is the amplitude of the process.

To decompose the cross section onto the definite ($q\bar{q}$) spin ($J^P$) 
 state one can integrate over $d^2k_t$ and write 
\begin{equation}
\frac{d\sigma(\gamma^*p\to(q\bar{q}p)}{dtdM^2}=4\pi e^2_qN_c
\sum_\lambda\int\frac{dz}{16\pi^2}\frac{z(1-z)|\mbox{M}_\lambda|^2}{16\pi s^2}=
\sum_\lambda\frac{e^2_qN_c}{(16\pi)^2}\sum_{J,\mbox{m}}|C^J_{\mbox{m}}|^2
\end{equation}
 with the coefficients $C^J_{\mbox{m}}$ 
given by the projection of the amplitude 
 with the help of the conventional spin rotation matrices
$d^J_{\mbox{m,m}'}(\theta)$
\begin{equation}
C^J_{\mbox{m}}=\int^1_{-1}d^J_{\mbox{m,m}'}(\theta)\frac{\mbox{M}_\lambda}{s}
\sqrt{z(1-z)}\sqrt{2J+1}d\cos\theta\; .
\end{equation}
The factor $\sqrt{z(1-z)}$ comes from the denominator in the $\delta$-function 
after the $dk^2_t$ integration of (2) while the $\sqrt{2J+1}$ reflects the 
normalization of the spin rotation matrices
$$\int^1_{-1}|d^J_{\mbox{m,m}'}(\theta)|^2d\cos\theta=\frac 2{2J+1}\; ,$$
$\mbox{m}$ and $\mbox{m}'$ are the projection of 
the spin $J$ onto the initial photon or quark axis respectively; the value of 
 $\mbox{m}$ is given by the helicity of heavy photon, while for the quark  
helicity $\lambda=\pm 1/2$ the value of $\mbox{m}'=(\lambda+\lambda')=\pm 1$ 
due to helicity conservation of the $\gamma\to q\bar{q}$-vertex; $\theta$ is 
the quark polar angle which may be expressed in terms of $z$ as 
$z=(1+\cos\theta)/2$.

The projection onto the isospin $I^G=1^+$ state gives just the factor $0.9$ 
in accordence with the well known $\omega\, :\,\rho\, =\, 1\, :\, 9$ ratio 
in Vector Dominance Model.

\section{Born amplitude for the open quark production}

\subsection{Gluon exchange}
Let us consider for the begining interaction with the quark target. 
The Born amplitude of the diffractive reaction
 $\gamma^*+q\to (q\bar{q}) +q$ is described by the 
sum of the diagrams  shown in fig.1.

At large energies $s=(q+p)^2\gg |q^2|+m^2_\rho$ the main 
contribution comes from the 
longitudinal polarizations of t-channel gluons ($l$ and $l+Q_{\mbox{tr}}$ 
in fig.1), i.e.  the spin part of gluon propagator
is given by
$$g_{\rho\sigma}\; =\; g^\perp_{\rho\sigma}\; +\; \frac{p'_\rho q'_\sigma + 
q'_\rho p'_\sigma}{(p'q')}\;\simeq\; \frac{p'_\rho q'_\sigma}{(p'q')}$$
with
$$q_\mu=q'_\mu+\frac{q^2}s p'_\mu;\; 
p_\mu=p'_\mu+\frac{m^2_N}sq'_\mu\simeq p_\mu;\; s=2(p'q');\; p'^2=q'^2=0\; .$$
Here $m_N$ is the target (quark or nucleon) mass, 
 $p$, $p''$ and $q$ are the 4-momenta of 
the target (initial quark or proton), recoil quark (or proton) and photon 
 correspondingly and we assume that the momentum transfered 
$Q^2_{\mbox{tr}}=(p-p'')^2$ is small. 
Indeed, in the forward direction the transverse 
component $Q_{\mbox{tr},t}=0$ and the longitudinal part gives $Q^2_{\mbox{tr}}
=t_{min}=-x^2 m^2_N\to 0$ for $x\ll 1$.

However the longitudinal t-channel (Coulomb-like) gluon looses the information 
 about the polarization of the target. Thus at least one gluon 
 must  have transverse polarization.
Indeed for the longitudinally polarized target with spin vector $s_\mu \| 
p_\mu$ the spin dependent part of the trace in the bottom of
 the diagram fig.1 (the target loop) looks like
\begin{equation}
B_{\sigma\sigma'}\; =\; \frac 12 
\mbox{Tr}[\hat{p}\gamma_5\gamma_\sigma(\hat{p}+\hat{l})\gamma_{\sigma'}]\; =\; 
-2i\epsilon^{\sigma\sigma'\alpha\beta}l_\alpha p_\beta
\end{equation}
where $\epsilon^{\sigma\sigma'\alpha\beta}$ is the antisymmetric tensor and 
 $\sigma$ ($\sigma'$) corresponds to the polarization of the left (right) 
t-channel gluon in fig.1. (Considering the forward amplitude we put the momentum
 transfer $Q_{\mbox{tr}}\simeq 0$.)\\
As $p_\mu$ is the longitudinal vector while the $l_\mu\simeq l_{t,\mu}$, two 
other indices $\sigma ,\sigma'$ should be the transverse (say, 
$g_{\rho\sigma}=g^\perp_{\rho\sigma}$) and the  longitudinal 
 $g_{\rho'\sigma'}\simeq\; \frac{p'_{\rho'} q'_{\sigma'}}{(p'q')}$ ones.
So the spin dependent contribution of the lower part of the
graph takes the form
\begin{equation}
g_{\rho\sigma} B_{\sigma\sigma'}g_{\rho'\sigma'}= B_{\rho\rho'}\simeq 
2(e^\perp_\rho\, p'_{\rho'}-p'_{\rho}e^\perp_{\rho'})/l^4.
\end{equation}
Here the gluon polarization vector $e^\perp_\nu=
i\epsilon_{\nu\mu}^\perp l^\mu$ (see Eq.(5)); 
$\epsilon_{\nu\mu}^\perp$ is the two-dimensional antisymmetric tensor acting in 
the transverse plane. The denominator $l^4$ comes from the two gluon 
($l$ and $l+Q_{\mbox{tr}}$) propagators (recall that in the forward direction 
$Q_{\mbox{tr}}\to 0$).

To find the contribution corresponding to the upper quark loop 
we use the formalism of ref.~\cite{BL,AHM} and calculate first the matrix 
element of the $\gamma\to q\bar{q}$ transition putting both the quarks (with 
helicities $\lambda$ and $\lambda'=\pm 1/2$) on mass shell.

For the transversely polarized photon the matrix element reads
\begin{equation}
\Psi_{\lambda ,\lambda'}(k'_t,z)=\frac 1{2\sqrt{z(1-z)}}
\bar{u}_\lambda(\gamma_\mu\cdot E^\pm_\mu)v_{\lambda'}=
\frac{\delta_{\lambda,\lambda'}}{2z(1-z)}(E^\pm k'_t)[2\lambda(1-2z)
\mp1],
\end{equation}
where $E^\pm=\frac 1{\sqrt{2}}(0,1,\pm i,0)$ is the photon polarization vector;
 the first factor $1/2\sqrt{z(1-z)}$ reflects the normalization ($1/\sqrt{2E_j};
 \;\; j=1,2$) of the quark fields.  

The  quark-gluon vertex conserves the helicity of the high energy quark and the 
momentum fraction $z$ is not changed either. Therefore only the transverse 
momentum of the quark $k'_t$ in Eq.(7) may be different from the final quark 
momentum $k_{jt}$. In order to simplify the projection onto the $J^P=1^-$ 
state we 
use the momentum $k_t$ in the rest frame of the $q\bar{q}$-system (the $z$ axis 
is directed along the target proton momentum) and put the light quark mass 
$m_q=0$. So the value of $k_{jt}=(M/2)\sin\theta$ (where $M=M_{q\bar{q}}$ 
 is the mass of the $q\bar{q}$ pair and $\theta$ is the quark polar angle). 

For the case of the longitudinal polarizations ($e^\|_\rho =p'_\rho$) 
 of both t-channel gluons the  quark-gluon vertex 
$\bar{u}(k')(\gamma_\mu\cdot p'_\mu)u(k'-l)=\left( (2k'-l)\cdot p'\right)$ 
 looks as the emission of a  
soft ($z_g\ll 1$)  gluon by a classical colour charge 
(i.e. by the 'current' $j_\mu=2k_{j\mu}$) 
 giving the factor $2(p'\cdot k_j)=z_j s$ 
for each vertex. This factor $z_j$ cancels the normalization factors 
($1/\sqrt{z_j}$) of the quark fields. 
So the contribution of the upper part of 
Feynman diagram shown in fig.1a takes the form~\cite{LMRT}
\begin{equation}
\Phi_\lambda=\frac{\delta_{\lambda,\lambda'}}{\bar{Q}^2+k'^2_t}(E^\pm k'_t)
[2\lambda(1-2z)\mp1]\;.
\end{equation}
The new factor $1/(\bar{Q}^2+k'^2_t)$ ( $\bar{Q}^2=z(1-z)Q^2$) comes from the 
energy denominator $\Delta 
E=E_{q\bar{q}}-E_{\gamma^*}=\frac{z(1-z)Q^2+k'^2_t}{z(1-z)}$ and corresponds 
to the propagator of quark $k'$. 

However in our case one of the t-channel gluons has the transverse 
polarization $e^\perp_\nu=i\epsilon_{\nu\mu}^\perp l^\mu$ (see Eq.(2)).
 When this gluon couples to the quark line it gives the
 factor $2(e^\perp\cdot k'_t)$ instead of the old one $2(p'\cdot k')=zs$ for 
 a longitudinal gluon polarization vector $e^\|_\mu=p'_\mu$. Note that, as 
it was confirmed by the explicite calculation of the Trace, within the LLA
 the quark-gluon vertex may be written as the emission of a soft ($z_g\ll 1$) 
 gluon by a classical colour charge 
(i.e. by the 'current' $j_\mu=2k_\mu$) for the polarized case also. 

Thus for the spin dependent part of the amplitute fig.1a in comparison with 
Eq.(8) one looses the factor $zs$ but gets the $4(e^\perp\cdot k'_t)$; an
extra factor of two reflects the contributions of two terms in the expression 
(6) (due to the permutation of gluon polarizations).

To select the leading logarithm in the $dl^2$ integral (coming from the region 
of $l^2\ll \bar{Q}^2$), we have to separate in 
the numerator the term proportional to $l^2_t$. There is no such  term in 
the fig.1a contribution, but in the case of fig.1b, where $k'=k_1+l\equiv k+l$,
 one can take the product $(E^\pm\cdot l_t)(e^\perp\cdot k_t)$ and put 
$k'\simeq k$ in the denominator.
\footnote{ The $(k_t\cdot l_t)$ term coming from 
the expansion of the denominator $\bar{Q}^2+(k+l)^2_t$ does not contribute as, 
after the averaging over the $\vec{l}_t$ direction (in the azimuthal plane), the 
product $(e^\perp\cdot k'_t)(k_t\cdot l_t)$ gives zero. Indeed, the intergral 
over the azimuthal angle $\phi$ reads  
$\int (k_\mu\cdot\epsilon^{\mu\nu}_\perp l_\nu)(k_\beta\cdot l_\beta)d\phi =0$.}

So the amplitude corresponding to the fig.1b takes the form
\begin{equation}
\mbox{M}^b_\lambda = \frac 29 4\pi s
\int\frac{dl^2}{l^4}\alpha_s^2\frac{[2\lambda
(1-2z)\mp 1]}{\bar{Q}^2+k^2_t}\frac{2(E^\pm\cdot l_t)(e^\perp\cdot k_t)}{zs}
\end{equation}
where $\frac 29$ is the colour coefficient.

Adding the contribution of another graph where the left t-channel gluon couples
 to the antiquark with $z_2=1-z$ (instead of $z_1=z$ for the quark) and 
neglecting the $k^2_t<M^2/4\ll\bar{Q}^2$ in our small ($M^2\sim m^2_\rho$) mass 
 region one finally gets
\begin{equation}
\frac{\mbox{M}_\lambda}s = \frac 29 4\pi
\int\frac{dl^2}{l^2}\alpha_s^2\frac{2\lambda
(1-2z)\mp 1}{\bar{Q}^4}x(E^\pm_\mu\cdot i\epsilon^{\mu\nu}_\perp k_t)
\end{equation}
Here we average over the dirrection of vector $l_t$ in azimuthal plane and 
the equalities $1/z+1/(1-z)=1/z(1-z)$ and $\frac 
1{z(1-z)s}=\frac{Q^2}{s\bar{Q}^2}=x/\bar{Q}^2$ (with $x=Q^2/s$) is used.

\subsection{Quark exchange}
Note that for the unpolarized case the amplitude given by the gluon exchange 
was proportional to the gluon density $xG(x,Q^2)$ and  in the Born 
approximation it tends to $const$ at $x\to 0$. The quark exchange corresponds
 (in the same Born approximation) to the quark structure function which was 
negligible ($xq(x,Q^2)\to 0$) at small $x\to 0$.

In contrast, the spin dependent part of the amlitude (10) is proportional to 
$x$, reflecting the fact that the polarized gluon distribution $x\Delta 
G(x,Q^2)\propto x$ (modular the $\alpha_s$ corrections) in the small $x$ 
region. The quark distribution $x\Delta q(x,Q^2)\propto x$ possesses
 the same property. Indeed, in the polarized DGLAP evolution  the splitting 
kernels $\Delta P_{GG}(z)$, $\Delta P_{Gq}(z)$ and $\Delta P_{qq}(z)$ have no 
$1/z$ singularity (such a singular term in $P_{GG}(z)\simeq 2N_c/z$ 
provides the $1/x$ behaviour of the unpolarized gluons $G(x,Q^2)\propto 1/x$).

Therefore now we can not omit the quark exchange amplitude.
 In other words after the $\gamma\to q\bar{q}$ transition one has consider not 
only the quark-quark scattering (shown in fig.1) but the antiquark-quark 
annihilation (see fig.2) as well. Thanks to the helicity conservation, in 
the forward direction ($Q_{\mbox{tr},t}=0$) the amplitude shown in fig.2 
conserves the helicity of initial quark line \cite{GGFL,KL}; it contains the 
factor $\delta_{\lambda_i,\lambda_f}\delta_{\lambda_i,\lambda_t}$, where:
 $\lambda_i=\lambda'$ and $\lambda_f$ are the helicities of the initial and 
final antiquark whilst  $\lambda_t$ denotes the helicity of target quark.\\ 
 The momentum fraction $z$ of the fast antiquark is also not changed . 
From this point of view both the 
gluon and the quark exchange act in the same way. The only difference is the 
 precise  form of the quark-quark amplitude.

In the small $x$ region, the leading logarithmic contribution of the 
quark-antiquark annihilation amplitude is well known \cite{GGFL,KL}.
\begin{equation}
T=2\pi\frac{C_F^2}{N_c}\int\alpha_s^2\frac{dl^2}{l^2}
\end{equation}
(with  $C_F=\frac{N_c^2-1}{2N_c}$).

Thus the contribution of the Feynman diagram shown in fig.2 
(fast antiquark with the 
 momentum fraction $z_2=1-z$ and $k_{2t}=-k_t$ 
  annihilates with the target quark) reads:
\begin{equation}
\frac{\mbox{M}^q_\lambda}s = 
-\delta_{\lambda,\lambda'}\delta_{\lambda',\lambda_t}\frac 89 \pi^2 
\int\frac{dl^2}{l^2}\alpha_s\left(\frac{C_F\alpha_s}{\pi}\right)
\frac{2\lambda(1-2z)\mp 1}{\bar{Q}^2}\frac{(E^\pm\cdot k_t)}{(1-z)s}\; .
\end{equation}

With the help of the relation $1/((1-z)s)=zx/\bar{Q}^2$ it may be written as
\begin{equation}
\frac{\mbox{M}^q_\lambda}s = 
-\delta_{\lambda,\lambda'}\delta_{\lambda',\lambda_t}\frac {16}9 \pi^2 
\int\frac{dl^2}{l^2}\alpha_s\left(\frac{C_F\alpha_s}{2\pi}\right)
\frac{2\lambda(1-2z)\mp 1}{\bar{Q}^4}zx(E^\pm\cdot k_t)\; .
\end{equation}

\section{Spin-spin asymmetry}
\subsection{Leading logarithmic corrections}
In order to make the (Born) calculation more realistic we have to include the
 'ladder evolution' gluons (shown symbolically by the dashed lines in fig.1,2)
 and to consider the process at the proton, rather than the quark, level; i.e.
 the heavy photon diffractive dissociation on the proton target. This is 
achieved by the replacements:
\begin{equation}
C_F\int\frac{\alpha_s}\pi \frac{dl^2}{l^2}\,\rightarrow\,\Delta 
G(x,\bar{Q}^2)\, ,
\end{equation}
for the gluon exgange amplitude and
\begin{equation}
C_F\int\frac{\alpha_s}{2\pi} \frac{dl^2}{l^2}\,\rightarrow\,\Delta 
q(x,\bar{Q}^2)
\end{equation}
for the antiquark-target quark annihilation.

Indeed, the logarithmic $dl^2/l^2$ integration in Eq. (10) is  nothing else 
 but the first step of the DGLAP
 evolution of the spin dependent gluon distribution
$\Delta G(x,\bar{q}^2)$ with the splitting function \cite{AP} $\Delta 
P_{Gq}=C_F\frac{1-(1-z)^2}z\simeq 2C_F$ for $z\ll 1$.  
Correspondingly, in the quark case of Eq. (13) this 
 is the first step of the DGLAP
 evolution of the spin dependent quark distribution
$\Delta q(x,\bar{q}^2)$ with the splitting function \cite{AP} $\Delta 
P_{qq}=C_F\frac{1+z^2}{1-z}\simeq C_F$ for $z\ll 1$.

Strictly speaking, even at zero transverse momentum $Q_{\mbox{tr},t}=0$ 
 one does not obtain the exact gluon  structure 
function, as a non-zero component of the longitudinal  
momentum is transferred through the two-gluon (two-quark) ladder. 
However, in the region of 
interest, $x\ll 1$, the value of $|t_{min}|=m^2_Nx^2$ is so small that we may 
safely put $t\equiv Q^2_{\mbox{tr}}=0$ and identify the ladder coupling to the 
 proton with $\Delta G(x,\bar{Q}^2)$ (or with $\Delta q(x,\bar{Q}^2)$ in the 
quark case); see \cite{LMRR} for more details. The arguments of the
 parton  structure 
function should be $x=(|Q^2|+M^2)/s$ and $\bar{Q}^2$.

Now in the leading log$(Q^2)$ approach the spin dependent parts of the 
 amplitudes (10,13)  read
\begin{equation}
\frac{\mbox{M}^G_\lambda}s = \frac 23 \pi^2
\alpha_s x\Delta G(x,\bar{Q}^2)\frac{2\lambda
(1-2z)\mp 1}{\bar{Q}^4}(E^\pm_\mu\cdot i\epsilon^{\mu\nu}_\perp k_t)\; ,
\end{equation}
\begin{equation}
\frac{\mbox{M}^q_\lambda}s = -\frac {16}9 \pi^2 
\alpha_s xq_\lambda(x,\bar{Q}^2)
\frac{2\lambda(1-2z)\mp 1}{\bar{Q}^4}z(E^\pm\cdot k_t)\; .
\end{equation}
and
\begin{equation}
\frac{\mbox{M}^{\bar{q}}_\lambda}s = \frac {16}9 \pi^2 
\alpha_s x\bar{q}_\lambda(x,\bar{Q}^2)
\frac{2\lambda(1-2z)\mp 1}{\bar{Q}^4}(1-z)(E^\pm\cdot k_t)
\end{equation}
for the antiquark component of the target wave function.\\
In the last two equations (17,18) the helicity of fast quark (antiquark) is 
 strongly correlated with the target helicity ($\lambda=\lambda_{target}$ 
 see Eqs. (12,13)) 
 and the polarized quark distribution $\Delta q =
q_{\lambda=1/2}-q_{\lambda=-1/2}$. On the other hand in Eq. (16) there is no 
such a correlation and calculating the cross section with fixed helicities of 
the initial photon and proton one has to sum two contributions ( with 
$\lambda=1/2$ and $\lambda=1/2$). 

\subsection{Asymmetry for diffractive open quark production}

In the same notation the forward amplitude for the unpolarized case 
 (and a transversely polarized photon) looks like:
\begin{equation}
\frac{\mbox{M}^{sym}_\lambda}s = 4\frac 23 \pi^2
\alpha_s xG(x,\bar{Q}^2)\frac{2\lambda
(1-2z)\mp 1}{\bar{Q}^4}(E^\pm_\cdot k_t)\; .
\end{equation}
In other words the whole amplitude is proportional to
$$\mbox{M}_\lambda\;\propto\; 
i \left(G(x,\bar{Q}^2)E^\pm_\mu g^\perp_{\mu\nu}+\frac 14
\Delta G(x,\bar{Q}^2)\cdot E^\pm_\mu i\epsilon^\perp_{\mu\nu}\right.$$
\begin{equation}
\left.
- z\frac 23 q_\lambda(x,\bar{Q}^2)E^\pm_\mu g^\perp_{\mu\nu}
+ (1-z)\frac 23 \bar{q}_\lambda(x,\bar{Q}^2)E^\pm_\mu g^\perp_{\mu\nu}\right) 
k_{t\nu}\; .
\end{equation}
Note that $E^\pm_\mu i\epsilon^\perp_{\mu\nu}=\pm E^\pm_\nu$. Therefore the 
 $\Delta G(x,\bar{Q}^2)$ term   changes the sign when one reverses 
the helicity of incoming photon.
\footnote{On the other hand changing the helicity of the target we get 
$-i\epsilon^\perp_{\mu\nu}$ instead of $+i\epsilon^\perp_{\mu\nu}$ (see Eq. 
(5))}.
For the quark case the amplidudes (17,18) change the sign either due to the
sign of $\Delta q=q^\uparrow -q^\downarrow$ or due to the second ($\mp 1$) term 
in the numerator corresponding to the photon helicity.
 
In terms of the spin-spin asymmetry $A=A_{LL}$, eq.(20) means that for the
heavy photon "elastic" diffractive dissociation 
  $\gamma^*+p\to (q\bar{q}) +p$ \footnote
{The polarization of the photon emitted by 
the 100\% polarized initial lepton in DIS is 
$P_{ph}=\frac{1-(1-y)^2}{1+(1+y)^2}$, where $y$ is the lepton momentum fraction
 carried by the photon in the target proton rest frame.}
\begin{equation}
A\;=\;\frac{\sigma^{\uparrow\uparrow}-\sigma^{\uparrow\downarrow}}
{\sigma^{\uparrow\downarrow}+\sigma^{\uparrow\uparrow}}\;=\;
2\frac{\frac 14\Delta G(x,\bar{Q}^2)G(x,\bar{Q}^2)}{(G(x,\bar{Q}^2))^2+
(\frac 14 \Delta G(x,\bar{Q}^2))^2}\;\simeq\; 
\frac{\Delta G(x,\bar{Q}^2)}{2G(x,\bar{Q}^2)}
\end{equation}
The arrows indicate the helicities of the incoming photon and the target 
 nucleon  and for the moment we omit the quark contribution in Eq. (21).

As in the case of $J/\Psi$ leptoproduction an extra factor 2 in Eq. (21) 
comes in because the cross section of the
diffractive process is proportional to the parton (gluon) density squared. 
Note, however, that  due to the 'Fermi motion' which washes out the effect, 
 the effective analysing power of the $\rho$-meson production turns out to 
be 4 times smaller in comparison with the nonrelativistic $J/\Psi$-meson (see 
\cite{JGN}). 

The whole expression in the limit of $G(x,\bar{Q}^2)\gg \Delta 
G(x,\bar{Q}^2)$ and $G(x,\bar{Q}^2)\gg
\Delta q(x,\bar{Q}^2),\;\Delta \bar{q}(x,\bar{Q}^2)$ (which is reasonable at 
small $x$) takes the form  
\begin{equation}
A_{\gamma^*\to q\bar{q}}\;\simeq
\frac{\Delta G(x,\bar{Q}^2)}{2G(x,\bar{Q}^2)}+\frac{2[z^2-(1-z)^2]
\left((1-z)\Delta \bar{q}(x,\bar{Q}^2)-z\Delta q(x,\bar{Q}^2)\right) }
{3[z^2+(1-z)^2]G(x,\bar{Q}^2)}
\end{equation}

It is anticipated
that the main part of the corrections (NLO 
contributions) are cancelled 
when calculating the asymmetry by
Eqs. (21,22). For example, the uncertainty coming from the value of 
the QCD coupling constant (more exactly from the scale at which $\alpha_s$ is 
 evaluated) do cancel in the ratio Eq.(22). Thus the accuracy of the 
 expressions (21,22) for the asymmetry 
is expected to
be even better than that of the unpolarized
 diffractive amplitude.

\section{Discussion}

In order to obtain the spin-spin asymmetry for the case of diffractive 
$\rho$-meson production we have to project the amplitudes (16-18) onto
 the spin $J^P=1^-$ state as it was discribed in sect.2 (see Eq. (4)). 

Of course at the begining the photon produces the $q\bar{q}$ pair in a pure 
 $J^P=1^-$ state with the $\theta$ distribution 
$(1\pm\cos \theta)/2=d^1_{1,\pm 1}(\theta)$. Indeed, the $\theta$ 
distribution is given by the factor 
$[2\lambda(1-2z)\mp1]$ in the matrix element (7) and for, say, photon helicity 
+1 ($E^+$) and $\lambda=1/2$ it is equal to $-2z=-(1+\cos\theta)$.
 However, interaction with the target distorts the $q\bar{q}$ state due to the
 $Q^2$ dependence of the structure functions and an extra factor $\bar{Q}^2=
z(1-z)Q^2=\frac{\sin^2\theta}4 Q^2$ in the denominators of Eqs. (16-18).

Strictly speaking the integral over $\cos\theta$ (4) should be done numerically 
 with the concrete parton distributions (in the amplitudes $\mbox{M}_\lambda$).
  Instead of this let us discuss the singularities and the main structure of 
the integral (4).

In the scaling limit (parton densities have no $Q^2$ dependence) the 
singularities at $\cos\theta=\pm1$ (i.e. $z\to 0$ or $z\to 1$) comes from the
$z^2(1-z)^2$ factor in the denominators $\bar{Q}^4$ of Eq.s (16-18). Note that
at fixed  mass $M_{q\bar{q}}=M$ the value of 
transverse momentum $k_t=\frac M2\sin\theta=\sqrt{z(1-z)}M$ and after the 
$J^P=1^-$ projection the integral (4) has 
only the logarithmic singularity (in the small $\bar{Q}^2=z(1-z)Q^2$ region) 
 which is usual for conventional deep inelastic process. 

Any small anomalous dimension $\gamma >0$ which reflects the $q^2$ behaviour of 
the partons (say, $G(x,q^2)\sim (q^2)^\gamma$ ) is enough to provide the 
convergence of the integral.

If, for simplicity, we assume that in the whole essential region of $q^2$ 
the small $x$ parton distribution may be parametrized as 
$f(x,q^2)=f(x,q^2_0)(q^2/q^2_0)^\gamma$ with the fixed value of $\gamma=const$,
 it is easy to calculate the coefficients $C^J_{\mbox{m}}$ using the 
identity
\begin{equation}
\int^\pi_0\sin^p\theta\,  d \theta\; =\; \sqrt{\pi}\frac{\Gamma(\frac 12 +\frac 
12 p)}{\Gamma(1 +\frac 12 p)}
\end{equation}
or for the case of small $\gamma\ll 1$ just to put
$$\int^1_0 \frac{dz}{z^{1-\gamma}}=\frac 1\gamma$$ 
Note that due to the factors $z$ or $1-z$ in the quark exchange amplitudes 
(17,18), we have only one singular point: $z\to 1$ at $\lambda =\frac 
12\lambda_{photon}$ for the 
 case of fast antiquark-quark annihilation (17) and $z\to 1$ at $\lambda 
=-\frac 12\lambda_{photon}$ for the case of Eq. (18). It leads to a 
 numerical suppression of the 
 quark exchange contribution in comparison  with the gluon exchange one where 
 for any $\lambda=\pm 1/2$ one has the singularity at both points $z\to 1$ and 
$z\to 0$.

Thus in the small $\gamma\ll 1$ limit the asymmetry for the diffractive $\rho$ 
production reads

\begin{equation}
A_{\gamma^*\to \rho}\;\simeq
\frac{\gamma(G)\Delta G(x,\bar{Q}^2)}{2\gamma(\Delta G)G(x,\bar{Q}^2)}
-\frac{\frac{\gamma(G)}{\gamma(\Delta \bar{q})}\Delta \bar{q}(x,\bar{Q}^2)
+ \frac{\gamma(G)}{\gamma(\Delta q)}\Delta q(x,\bar{Q}^2) }
{6G(x,\bar{Q}^2)}
\end{equation}
where $\gamma(f)$ denotes the anomalous dimension of the structure function 
$f=G,\, \Delta G,\; q,\, \bar{q},\, \Delta q,\, \Delta \bar{q}$.

Hopefully the expression (24) gives us some impression about the expected 
 value of 
asymmetry $A_{\gamma^*\to\rho}$ and may be used to estimate the effect in the 
future experiments.

\vspace {1cm}

{\bf Acknowledgements:} I would like to thank A. Borissov and T. Gehrmann  
for useful discussions.

The work was supported by the INTAS grant 93-0079-EXT and by the 
Volkswagen-Stiftung.

\vspace {1cm}
\centerline{{\bf Figure Captions}}
\vspace {0.5cm}
 
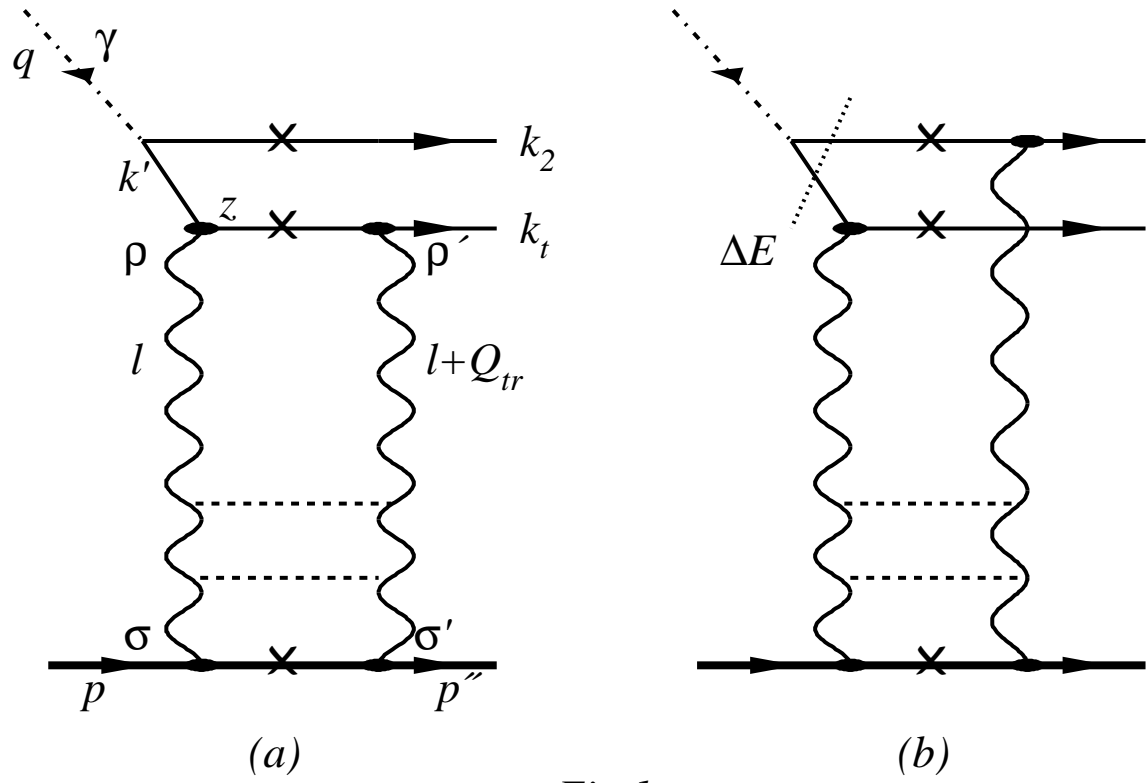
{\bf Fig. 1} Diffractive open $q\bar{q}$ production in high energy $\gamma^*p$ 
collisions via the two gluon exchange.

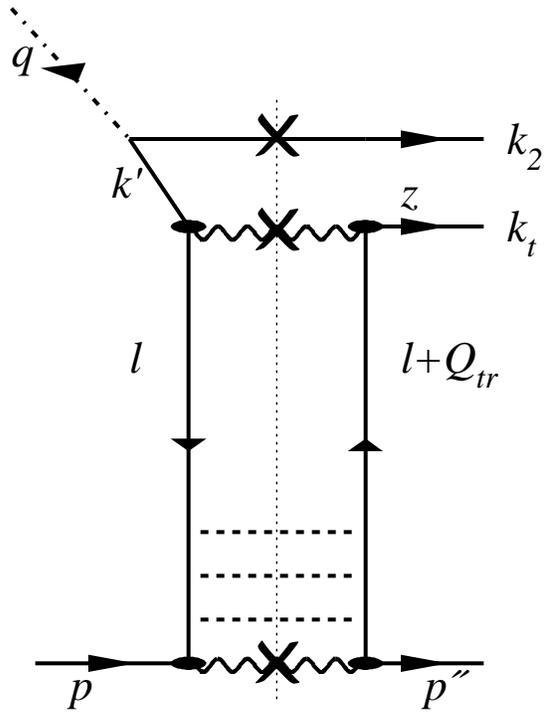
{\bf Fig. 2} Diffractive open $q\bar{q}$ production in high energy $\gamma^*p$ 
collisions via the quark exchange.

\end{document}